# Field Formulation of Parzen Data Analysis

D. Horn
Sackler School of Physics and Astronomy
Tel Aviv University, Tel Aviv, Israel

**Abstract**

The Parzen window density is a well-known technique, associating Gaussian kernels with data points. It is a very useful tool in data exploration, with particular importance for clustering schemes and image analysis. This method is presented here within a formalism containing scalar fields, such as the density function and its potential, and their corresponding gradients. The potential is derived from the density through the dependence of the latter on the common scale parameter of all Gaussian kernels. The loci of extrema of the density and potential scalar fields are points of interest which obey a variation condition on a novel indicator function. They serve as focal points of clustering methods depending on maximization of the density, or minimization of the potential, accordingly. The mixed inter-dependencies of the different fields in d-dim data-space and 1-d scale-space, are discussed. They lead to a Schrődinger equation in d-dim, and to a diffusion equation in (d+1)-dim.

## 1. Introduction

Data analysis has become a dominant occupation of modern science; to the extent it is often referred to as belonging to Data Science. Physics has always relied on data collection and analysis, and some of the modern tools of machine learning were motivated by physical intuition. We try to build on such intuition in this analysis of the Parzen-window distribution [1], an important tool which has been introduced in 1965, and still serves the goal of pattern recognition [2]. It has recently been shown [3] to allow for a natural decomposition into Weight and Shape components, where the former entails semi-global features and the latter depends on local features of the Parzen distribution. In particular, Shape is the exponential of the potential *V* that has been introduced in Quantum Clustering [4].

Although the formalism is developed for statistical data analysis, the Weight-Shape decomposition [3] makes use of an analogy with statistical information theory [5]. Here we present a novel formulation based on vector and scalar fields and establish their inter-relations both in data-space and in 1-d scale-space. The mixed behavior of the different variables leads to a Schrődinger equation in d-dim data space, and to a diffusion equation in the (d+1)-dim scale and data space. A special role is played by the potential function which is derived from the Parzen distribution function through its dependence on the scale parameter.

## 2. The Parzen probability distribution and its potential function.

Consider the case of data points within some Euclidean data - or feature - space of dimension d, with possible positive attributes (e.g. intensities) $I_i$, described by an experimental distribution

$$Q(\mathbf{y}) = \frac{1}{N}\sum_i I_i \delta(\mathbf{y} - \boldsymbol{\mu}_i) \tag{1}$$

where $N = \sum_i I_i$. In doing so, we assume that the measurement errors are small by comparison to the distances between the points and to the resolution with which we wish to study the data (otherwise replace the delta-functions by Gaussians with appropriate widths). In order to investigate the behavior of the data we make use of the Parzen-window [1] probability distribution

$$\psi_q(\mathbf{x}) = \int d\mathbf{y}\, Q(\mathbf{y})\, K_q(\mathbf{x} - \mathbf{y}) \tag{2}$$

where we employ the Gaussian kernel

$$K_q(\mathbf{x} - \mathbf{y}) = \exp(-q\,(\mathbf{x} - \mathbf{y})^2). \tag{3}$$

$q = \frac{1}{2\sigma^2}$ is the resolution, with $\sigma$ being the Gaussian width.

$\psi_q(\mathbf{x})$ is the Weierstrass transform of $Q(\mathbf{y})$. As such, it may be defined for any probability distribution and it obeys

$$\int d\mathbf{z}\, K_{q_2}(\mathbf{x} - \mathbf{z})\, \psi_{q_1}(\mathbf{z}) = \left(\frac{\pi}{q_1+q_2}\right)^{d/2} \psi_{q_3}(\mathbf{x}) \quad \text{with} \quad q_3 = \frac{q_1 q_2}{q_1+q_2} \,. \quad (4)$$

Following [3] we define a relative probability weight

$$p_q(\mathbf{x}|\mathbf{y}) = \frac{K_q(\mathbf{x}-\mathbf{y})}{\psi_q(\mathbf{x})}, \quad (5)$$

which represents the influence of point $\mathbf{y}$ in $Q(\mathbf{y})$ on point $\mathbf{x}$ in $\psi(\mathbf{x})$, and is properly normalized at any point $\mathbf{x}$ through

$$\int d\mathbf{y}\, Q(\mathbf{y})\, p_q(\mathbf{x}|\mathbf{y}) = 1 \,. \quad (6)$$

Let us introduce the potential function $V_q(\mathbf{x})$ through

$$V_q(\mathbf{x}) = -q \frac{\partial}{\partial q} \log \psi_q(\mathbf{x}) = q \int d\mathbf{y}\, Q(\mathbf{y}) (\mathbf{x}-\mathbf{y})^2 p_q(\mathbf{x}|\mathbf{y}) \equiv q <(\mathbf{x}-\mathbf{y})^2> \quad (7)$$

where the last equality serves as the definition of an expectation value under the probability defined by Eqs. (5) and (6). It follows then that

$$\psi_q(\mathbf{x}) = \exp\left(-\int_0^q \frac{V_p(\mathbf{x})}{p} dp\right), \quad (8)$$

demonstrating that the potential $V_q(\mathbf{x})$ contains the information of the Parzen distribution function through its dependence on the scale parameter $q$. This implies that

$$\psi_{q_2}(\mathbf{x}) = \psi_{q_1}(\mathbf{x}) \exp\left(-\int_{q_1}^{q_2} \frac{V_p(\mathbf{x})}{p} dp\right)$$

for $q_1 < q_2$, relating low-resolution to high-resolution, in contradistinction to Eq. (4) which describes straight-forward relations of high-resolution to low resolution.

Based on Eqs. (2) and (7) one can derive [3] the following differential equation

$$-\frac{1}{4q} \nabla^2 \psi_q + V_q \psi_q = \frac{d}{2} \psi_q \quad (9)$$

which is a Schrödinger equation obeyed by $\psi$ with the potential $V_q(\mathbf{x})$ in a d-dimensional Euclidean space. This coincides with the formalism of Quantum Clustering (QC) [4] without invoking any quantum mechanical interpretation.

### 3. Entropy and the Weight-Shape decomposition.

Following [3], we define an entropy function at point $\mathbf{x}$

$$H_q(\mathbf{x}) = -\int d\mathbf{y}\, Q(\mathbf{y})\, p_q(\mathbf{x}|\mathbf{y}) \log p_q(\mathbf{x}|\mathbf{y}) = -< \log p_q(\mathbf{x}|\mathbf{y}) > \quad (10)$$

and note that it can be rewritten as

$$H_q(\mathbf{x}) = \log \psi_q(\mathbf{x}) + V_q(\mathbf{x}) \quad (11)$$

Eq.(11) has an analog in statistical mechanics [5], where $H$ is the entropy, $V$ is the average energy of a canonical ensemble, and $\psi$ is its partition function. Also Eq. (7) has an analog in statistical mechanics. The partition function of the canonical ensemble is $Z = \sum_i e^{-\beta E_i}$ and its average energy is

$\sum_i E_i e^{-\beta E_i} = -\frac{\partial}{\partial \beta} \log Z$. Replacing $Z$ by $\psi$ and $\beta$ by $q$, we find again that the potential $V$ is analogous to average energy. Note, however, that the scalar fields $\psi$, $V$ and $H$ are functions of space, rather than constants of a physical system [3].

Employing the functions $W(\mathbf{x}) = e^{H(\mathbf{x})}$ and $S(\mathbf{x}) = e^{-V(\mathbf{x})}$ we rewrite Eq. (11) as

$$\psi_q(\mathbf{x}) = W_q(\mathbf{x}) S_q(\mathbf{x}), \tag{12}$$

which has been named [3] the Weight-Shape decomposition of $\psi(\mathbf{x})$. Since $V(\mathbf{x}) \geq 0$, it follows that $S(\mathbf{x}) \leq 1$.

## 4. Vector Fields

Let us introduce vector fields in order to deal with the locations of extrema of the scalar fields $\psi_q(\mathbf{x})$, $V_q(\mathbf{x})$ and $H_q(\mathbf{x})$. We define the vector field $\mathbf{D}$ as

$$\mathbf{D} = -\nabla \log \psi_q(\mathbf{x}) = 2q < \mathbf{x} - \mathbf{y} > \tag{13}$$

Applying the grad operator to the potential we find (see Appendix)

$$\nabla V_q(\mathbf{x}) = \mathbf{D} + V\mathbf{D} - \mathbf{E} \quad \text{where} \quad \mathbf{E} = 2q^2 < (\mathbf{x} - \mathbf{y})^3 > . \tag{14}$$

Finally, we find that the remaining interesting vector, the gradient of entropy, obeys

$$\nabla H_q(\mathbf{x}) = \nabla V_q(\mathbf{x}) + \nabla \log \psi_q(\mathbf{x}) = V\mathbf{D} - \mathbf{E}. \tag{15}$$

The two new vector fields allow us to exhibit the different locations of the three extrema of the scalar fields. Local maxima (and minima) of the probability function occur at $\mathbf{D}=0$, maxima (and minima) of entropy occur where $\mathbf{E} = V\mathbf{D}$ and minima (and maxima) of the potential (or maxima and minima of Shape) occur where $\mathbf{E} = V\mathbf{D} + \mathbf{D}$. All extrema coincide if both $\mathbf{D}=0$ and $\mathbf{E}=0$ at the same value of $\mathbf{x}$. Otherwise, one finds at maxima of the Parzen probability that $\nabla V_q(\mathbf{x}) = -\mathbf{E}$.

From Eq. (13) we conclude that

$$q \frac{\partial}{\partial q} \mathbf{D} = \mathbf{D} + V\mathbf{D} - \mathbf{E} = \nabla V \tag{16}$$

which implies that maxima of $\psi_q$, where $\mathbf{D} = 0$, change with $q$ in a direction determined by $-\nabla V$, i.e. toward close-by minima of $V$. Eq. (16) leads to a new interpretation of QC: clustering based on minimization of $V$ may be interpreted as clustering based on *stationarity of $\mathbf{D}$ with respect to changes in the scale-parameter q.*

## 5. Extrema of Scalar Fields

Examples of the behavior of $\log \psi$ and of $V$ are demonstrated below for a data set of 9000 observed galaxies (with red-shift in the domain $0.47 \pm 0.005$) regarded as points in spherical angles $\theta$ and $\varphi$ within some limited range. Whereas for $\sigma = \frac{1}{\sqrt{2q}} = 2$ (in angle degree units) the two fields exhibit many extrema, there exist clear differences for larger sigma, e.g. $\sigma=10$, where $\log \psi$ has one maximum whereas $V$ maintains several minima.

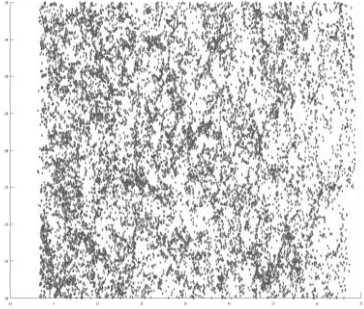

A.  Loci of 9000 Galaxies within some range of spherical angles.

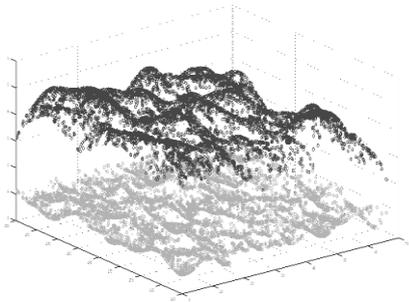

B. $\log \psi$ (top) and $V$ (bottom) displayed over the data plane A, using σ=2.

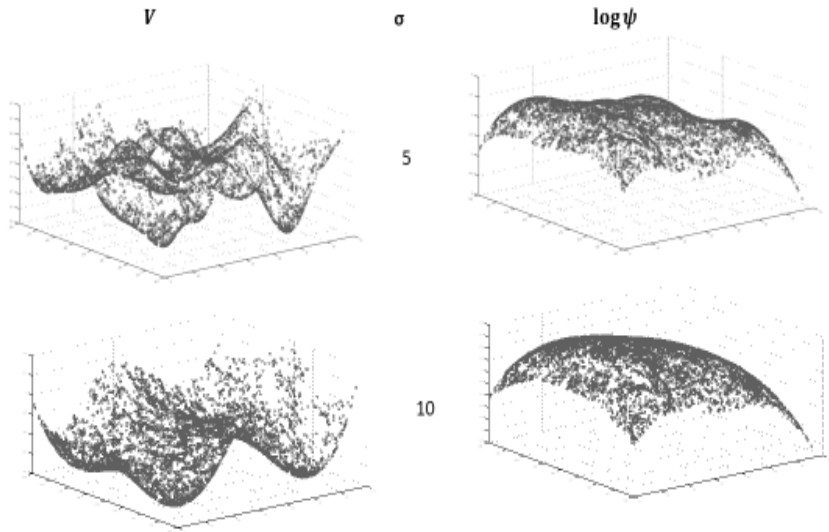

C.  Surfaces of V and $\log \psi$ for increased values of the Gaussian widths.

Fig.1. Demonstration of different structures of $\log \psi$ and of $V$ for a set of galaxies downloaded from the Sloan Digital Sky Server DR12.

The different extrema can be used for clustering formation. Clustering methods based on probability maximization are known as Mean Shift algorithms [6] [7]. QC [4] is a clustering method based on potential minimization. They were generalized in [3] into three families of algorithms, together with another method based on entropy maximization. A hierarchical structure of *V* at various scales for any complex data problem, can be obtained by employing Hierarchical Maximum Shape Clustering (HMSC), as defined in [3].

Information regarding the different extrema of $\psi$ and *V* can be united into one equation by considering another scalar function

$$U_q(\mathbf{x}) \equiv \mathbf{D}^2 = (2q <\mathbf{x}-\mathbf{y}>)^2 . \tag{17}$$

Using the relations spelled out above, it follows that the condition

$$q \frac{\partial}{\partial q} U_q(\mathbf{x}) = -2\nabla \log \psi_q(\mathbf{x}) \cdot \nabla V_q(\mathbf{x}) = 0 \tag{18}$$

implies that either $\psi$ or *V* (or both) reach their extrema at the **x** values for which Eq. (18) holds. *U* is non-negative, therefore *U*=0 is a minimum of *q*. It corresponds to extrema of $\psi$ which are associated with $\mathbf{D} = 0$. Other values of *U* which obey eq. (18) are associated with extrema of *V* which occur whenever $\frac{\partial}{\partial q}\mathbf{D}_q = 0$.

Eq. (18) is a concise statement regarding the points of interest in our data-analysis method. In analogy with statistics, one may view this equation as an inference method finding the correct parameter *q* which leads to points of interest at given values of **x**.

An interesting comparison can be made with the score function [8] in statistics, defined by

$$u(\theta) = \frac{\partial}{\partial \theta} \log f(\mathbf{x}|\theta)$$

for a probability function *f(***x***)* which depends on the parameter $\theta$. The average of *u* vanishes and its variance is known as Fisher's information [9]. The analogy of *f* with $\psi$ and *u* with *V* is incomplete, because the integral of $\psi$ is not normalized to 1. As a result, whereas the expectation value of *u* vanishes, the analogous statement in our analysis becomes

$$(\frac{q}{\pi})^{\frac{d}{2}} \int d\mathbf{x}\, \psi_q(\mathbf{x}) V_q(\mathbf{x}) = \frac{d}{2} . \tag{19}$$

Although all extrema may be regarded as points of interest, some are of more interest than others: extrema that remain fixed in **x** for a range of *q* values which is large compared with that of other points of interest. This criterion, introduced by Roberts [10], allows for searching for scales which correspond to important properties of the data. Thus it sub-serves the search for good clustering of the data [3,4,10].

## 6. Relation to the Diffusion Equation

The diffusion equation has served as the basis for various data analysis schemes such as the scale-space approach [11] and harmonic analysis expansion [12]. For the sake of completeness, we demonstrate in this section how it fits into our analysis.

The Gaussian function

$$G_t(\mathbf{x}) = \left(\frac{q}{\pi}\right)^{d/2} \exp(-q\,\mathbf{x}^2) = \left(\frac{q}{\pi}\right)^{d/2} K_q(\mathbf{x}) \qquad \text{with} \quad q = \frac{1}{4t} \qquad (20)$$

serves as the Green's function of the diffusion equation

$$\frac{\partial}{\partial t} G_t(\mathbf{x}) + \nabla^2 G_t(\mathbf{x}) = 0 \,. \qquad (21)$$

It follows that the solution to the diffusion equation with the boundary condition $\varphi_0(\mathbf{x}) = Q(\mathbf{x})$ at $t=0$ is provided by

$$(4t\pi)^{-d/2} \varphi_t(\mathbf{x}), \quad \text{where} \quad \varphi_t(\mathbf{x}) = \psi_q(\mathbf{x}) \quad \text{for} \quad q = \frac{1}{4t}. \qquad (22)$$

The factor $(4t\pi)^{-d/2}$ is needed to guarantee the correct limit at $t=0$. Interesting conclusions are the connections with the Schrödinger equation (9), the potential $V_q(\mathbf{x})$ of Eq. (7) and the Weight-Shape decomposition (12). Similarly, we obtain

$$v_t(\mathbf{x}) = V_q(\mathbf{x}) \quad \text{and} \quad h_t(\mathbf{x}) = H_q(\mathbf{x}) \quad \text{for} \quad q = \frac{1}{4t}. \qquad (23)$$

Let us discuss a few examples.

The first is $\left(\frac{q}{\pi}\right)^{d/2} \varphi_t(\mathbf{x}) = G_t(\mathbf{x})$ where
$v_t(\mathbf{x}) = (\mathbf{x})^2/4t$, $h_t(\mathbf{x}) = H_q(\mathbf{x}) = 0$, and $W=1$
which is to be expected from $Q = \delta(\mathbf{x})$.

The second is derived from of a one-dimensional step-function $Q = \theta(x)\theta(L-x)/L$.
This leads to
$$\psi(x) = \frac{1}{L}\int_0^L \exp\left(-\frac{(x-y)^2}{4t}\right) dy = \sqrt{\pi t}\left[\text{erf}\left(\frac{x}{\sqrt{4t}}\right) - \text{erf}\left(\frac{x-L}{\sqrt{4t}}\right)\right]. \qquad (24)$$

This function has been displayed as Fig. 2 in [3], together with its $W$ and $S$ components, demonstrating that Shape accentuates the edge effects of the distribution.

A third example is the plane wave solution of the diffusion equation:

$$\varphi_t(\mathbf{x}) = \exp(i\mathbf{k}\cdot\mathbf{x} - \mathbf{k}^2 t) \,.$$

It can be generated from $Q = \exp(i\mathbf{k}\cdot\mathbf{y})$, and it leads to $v_t(\mathbf{x}) = (\mathbf{x})^2$, $h_t(\mathbf{x}) = i\mathbf{k}\cdot\mathbf{x}$ and, correspondingly, $W = \exp(i\mathbf{k}\cdot\mathbf{x})$, $S = \exp(-\mathbf{k}^2 t)$. Here the WS decomposition exhibits a separation of variables, with $W$ and $S$ being separate $\mathbf{x}$ and $t$ dependent functions.

Replacing the plane wave by a wave packet, $Q = \exp(i\mathbf{k_0}\cdot\mathbf{y} - (\mathbf{y}\Delta)^2/2)$
corresponding to average momentum $\mathbf{k_0}$ and Gaussian widt $\Delta$, will result in

$$\varphi_t(\mathbf{x}) = \exp[(i\mathbf{k_0}\cdot\mathbf{x} - \mathbf{k_0}^2 t - 2t\Delta^2\mathbf{x}^2)/(1+2t\Delta^2)]$$

which no longer displays any separation of variables.

## 7. Conclusions

The three equations (7) (13) and (16) define the behavior and inter-relations of the potential and the Parzen probability in both **x** and $q = \frac{1}{2\sigma^2}$ :

- Maxima (and minima) of the probability $\psi_q(\mathbf{x})$ occur where **D** vanishes, i.e. when $<\mathbf{x}-\mathbf{y}> = 0$.
- Minima (and maxima) of the potential $V_q(\mathbf{x})$ occur when $\nabla <(\mathbf{x}-\mathbf{y})^2> = 0$, which is also when the variation of **D** with respect to $q$ vanishes, $\frac{\partial}{\partial q}\mathbf{D}_q(\mathbf{x}) = 0$.
- Maxima of the Parzen probability $\psi_q(\mathbf{x})$ and minima of the potential $V_q(\mathbf{x})$ will only seldom co-occur. This happens if both vectors **D** and **E** vanish, i.e. if both $<(\mathbf{x}-\mathbf{y})> = 0$ and $<(\mathbf{x}-\mathbf{y})^3> = 0$.
  The relation between these two types of extrema is reminiscent of the relation between phase-velocity $\omega/k$ and group-velocity $\partial\omega/\partial k$ of waves with frequency $\omega$ and wave-number $k$: whereas the maxima of $\psi_q(\mathbf{x})$ obey $\mathbf{D} = 0$, the minima of $V_q(\mathbf{x})$ obey $\partial \mathbf{D}/\partial q = 0$, which is a very relevant condition for interesting loci.

Different locations of point-minima of $V(\mathbf{x})$, for fixed $q$, are, by reasons of continuity, separated by maxima of $V(\mathbf{x})$, which form manifolds of dimension d-1. If a minimum $\nabla V(\mathbf{x}) = 0$ occurs along a 1-dim string, its separating manifold has dimension d-2. These separating manifolds serve to define clusters of data points. The latter can be calculated by manipulations of replica points, as in [4] and in [3]. Analogous statements hold for extrema of $\psi$ and of $H$, leading to different clustering methods [3].

We defined the points of interest to be the extrema of $\psi_q(\mathbf{x})$ and $V_q(\mathbf{x})$. All these extrema occur when the indicator function $U$ is stationary with respect to $q$,

$$\frac{\partial}{\partial q}U_q(\mathbf{x}) \equiv \frac{\partial}{\partial q}(2q<\mathbf{x}-\mathbf{y}>)^2 = 0.$$

In the investigation of data in the format of Eq. (1), particular attention should be given to extrema which remain stable for quite a large range of $q$. The latter should reveal important internal structure of these data. [10]

Points of interest may also accumulate along 1- or 2-dim structures in data-space. Such phenomena may exist in big data, where a panorama of hills (well-separated Gaussian structures) may be replaced by an alpine crest in data-space [13]. Eq. (24) is an example of a linear string that can serve as an attractor of QC in data-space [3].

## Appendix: Mathematical Relations

From the various definitions we can derive the following relations

$$q\frac{\partial}{\partial q}\frac{1}{\psi} = -\frac{1}{\psi}q\frac{\partial}{\partial q}\log\psi = \frac{V}{\psi} \qquad \nabla\frac{1}{\psi} = \frac{D}{\psi}$$

$$\nabla <f> = D<f> + <\nabla f> -2q<(x-y)f>$$

$$q\frac{\partial}{\partial q}<f> = V<f> + <q\frac{\partial}{\partial q}f> -q<(x-y)^2 f>$$

Thus

$$\nabla V_q(x) = 2q^2 <(x-y)^2><x-y> + 2q<x-y> - 2q^2<(x-y)^3>$$

which leads to Eq. (14).

Further derivatives of the potential are

$$q\frac{\partial}{\partial q}V = V + V^2 - M \qquad \text{where} \quad M = q<(x-y)^4>$$

i.e.

$$q\frac{\partial}{\partial q}V = q<(x-y)^2> + (q<(x-y)^2>)^2 - q<(x-y)^4>$$

and

$$q\frac{\partial}{\partial q}\nabla V = q\frac{\partial}{\partial q}(D + VD - E) = (D + VD - E)(1+V) + D(V+V^2 - M) - q\frac{\partial}{\partial q}E = (1+V)\nabla V + $$
$$+D(V+V^2-M) - (2E + VE - 2q^3<(x-y)^5>).$$

Note that

$$q^2\frac{\partial^2}{\partial q^2}D_q(x) = q\frac{\partial}{\partial q}\nabla V - \nabla V.$$

Near the potential minimum this becomes

$$q^2\frac{\partial^2}{\partial q^2}D_q(x) \approx D(V+V^2-M) - (2E + VE - 2q^3<(x-y)^5>)$$

**Low resolution limits.**
Expanding the Gaussian kernel in $q$, one may express $\psi_q(x)$ in terms of a Taylor series of weighted moments of the data,

$$\psi_q(x) = \sum_n \frac{(-q)^n}{n!} \int dy\, Q(y)\, (x-y)^{2n},$$

which, for the explicit data representation of Eq. (1) becomes

$$\psi_q(x) = \frac{1}{N}\sum_n \frac{(-q)^n}{n!}\sum_i I_i\, (x-\mu_i)^{2n} \quad .$$

In particular, in the asymptotic limit of low $q$ one finds

$$\psi_q(\mathbf{x}) \approx 1 - \frac{q}{N}\sum_i I_i \ (\mathbf{x} - \boldsymbol{\mu}_i)^2 \ .$$

The first term corresponds to the low limit of $W \approx 1 + O(q^2)$ and the second to the low limit of $S \approx 1 - V \approx 1 - q\sum_i I_i \ (\mathbf{x} - \boldsymbol{\mu}_i)^2/N + O(q^2)$ .

## Acknowledgments

I thank Isaac Meilijson and Zeev Schuss for helpful discussions, and Itay Fisher for his help with numerical data analysis. This research was partially funded by the Blavatnik Cyber Center of Tel Aviv University.